\documentclass[letter,
               biblatex,     
               ]{jacow}
\usepackage{pdfpages,multirow,ragged2e} %
\usepackage{svg}
\usepackage{listings}
\usepackage{float}      
\usepackage{placeins}   
\usepackage{authblk}
\makeatletter%
	\ifboolexpr{bool{xetex}}
	 {\renewcommand{\Gin@extensions}{.pdf,%
	                    .png,.jpg,.bmp,.pict,.tif,.psd,.mac,.sga,.tga,.gif,%
	                    .eps,.ps,%
	                    }}{}
\makeatother
\ifboolexpr{bool{xetex} or bool{luatex}} 
 {}                                      
 {\usepackage[utf8]{inputenc}}           

\usepackage[USenglish]{babel}
\usepackage{subcaption}

\ifboolexpr{bool{jacowbiblatex}}%
 {%
  \addbibresource{LLRF2025_LEMP.bib}
 }{}
\begin{document}

\title{Digital Low-Level RF system for the Linac Electronics Modernization Plan at LCLS}

\author[1]{N. Sawai}
\author[1]{J. Diaz Cruz}
\author[1]{A. Benwell}
\author[1]{S. Hoobler}
\author[2]{Q. Du}
\author[2]{S. Murthy}
\author[2]{L. Doolittle}
\affil[1]{SLAC National Accelerator Laboratory, Menlo Park, CA, USA}
\affil[2]{Lawrence Berkeley National Laboratory, Berkeley, CA, USA}

\maketitle

\begin{abstract}
The LCLS began operations in 2009, utilizing SLAC's normal-conducting (NC) LINAC, which features control equipment dating back to the 1960s and 1980s. The Linac Electronics Modernization Plan (LEMP) aims to replace the legacy control equipment with a system based on the open-source Marble carrier board and Zest+ digitizer board, both of which are used in the LCLS-II HE LLRF system. Adaptation of the LLRF system developed for the continuous-wave (CW) superconducting RF (SRF) LCLS-II\cite{Serrano:ICALEPCS2017-THSH202} to the short-RF pulse NC LCLS includes leveraging the knowledge and experience gained from recent LLRF projects at SLAC and efficiently reusing the core functionality of the hardware and code base developed for previous projects, in collaboration with LBNL, FNAL and JLAB. A prototype has been deployed and tested at station 26-3, demonstrating RF generation/control, interlocks, triggers, and waveform capture. Here, we describe the hardware, firmware and software infrastructure, highlight key features, and present initial test results.
\end{abstract}

\section{Introduction} 

The Linac Coherent Light Source (LCLS) at SLAC National Laboratory is a U.S Department of Energy user facility in operation since 2009. The LCLS is built upon one-third of the original SLAC normal-conducting (NC) copper accelerator designed and installed in the 1960s.  It uses approximately 80 high-power klystrons operating with a repetition rate of 120 pulses per second and with an RF pulse width of about $5 \mu s$. 

As of 2025, the accelerator control, including LLRF, reuses control equipment dating back to the 1960s and 1980s. The Linac Electronics Modernization Plan (LEMP) aims to update the aging infrastructure of the NC LINAC to overcome part obsolescence, dependence on staff who retain specialized knowledge for working on legacy equipment, and limits set by legacy equipment to improve system performance.

The LEMP LLRF design takes advantage of recent work on LCLS-II LLRF systems by incorporating key hardware, firmware, software and design concepts.  This greatly aligns the two accelerator systems, increases the speed of development of new features, and greatly reduces the complexity of maintenaning multiple accelerators.  The system is the product of major collaboration between several RF development teams including SLAC, LBNL, FNAL, and JLab during the LCLS-II project, and then continued collaboration between SLAC and LBNL during the initial phases of the LEMP effort. 

\begin{figure}[h]
  \centering
  \begin{subfigure}{\columnwidth}
    \centering
    \includegraphics[width=\columnwidth]{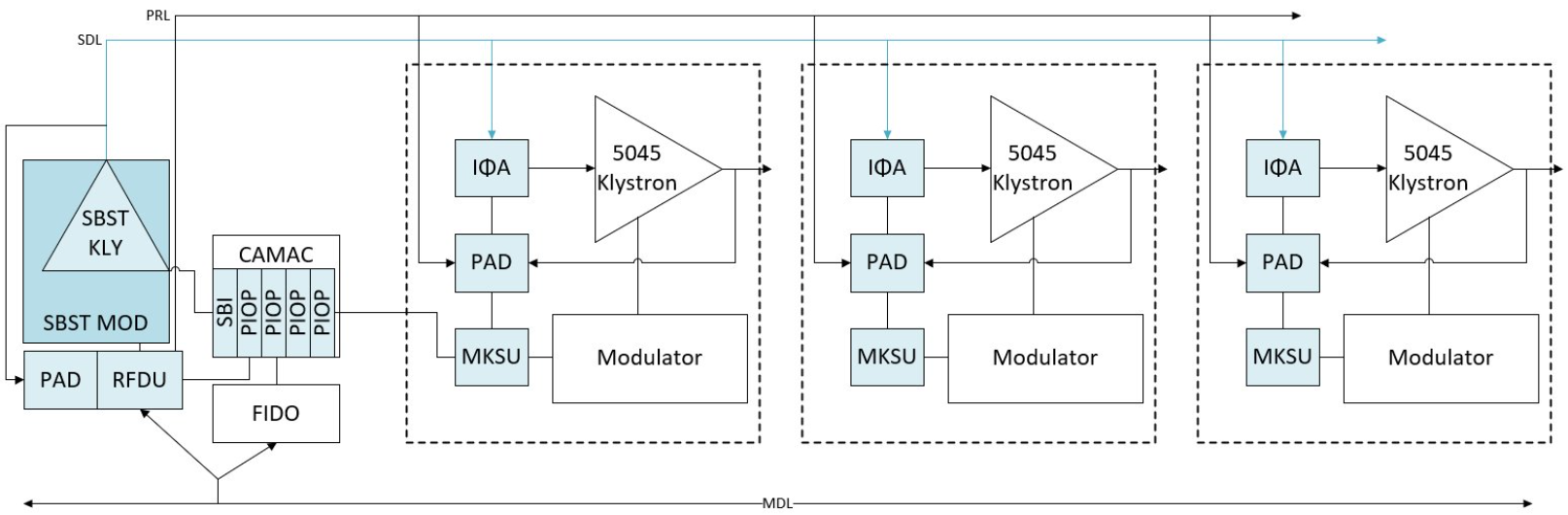}
    \caption{Legacy LCLS configuration.}
  \end{subfigure}

  \vspace{1em} 

  \begin{subfigure}{\columnwidth}
    \centering
    \includegraphics[width=\columnwidth]{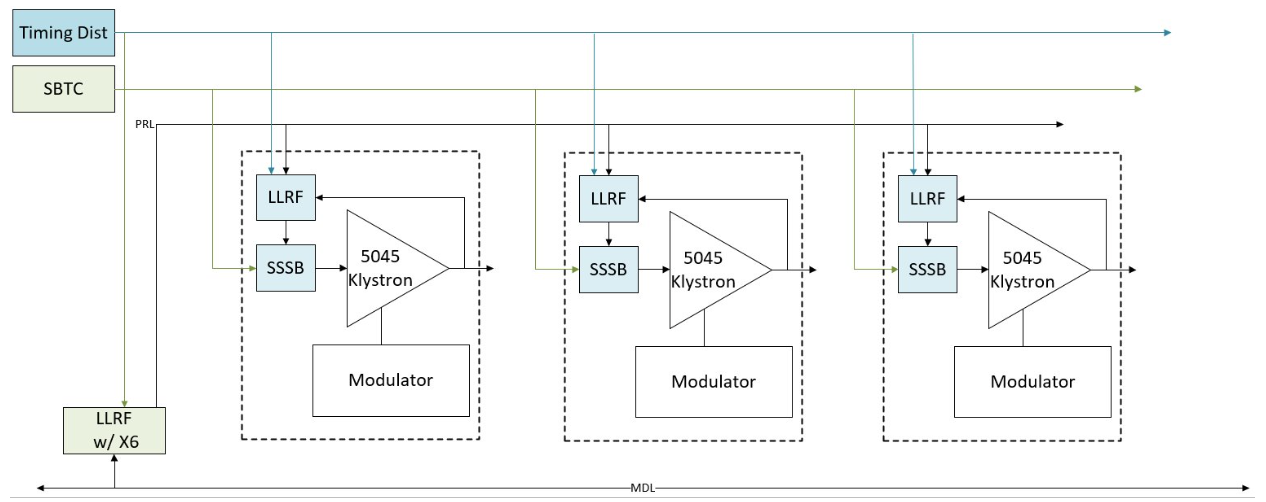}
    \caption{New LCLS LEMP configuration}
  \end{subfigure}

  \caption{LEMP sector upgrade.}
  \label{fig:LCLS_configuration}
\end{figure}

Fig.~\ref{fig:LCLS_configuration} shows the legacy LCLS configuration and the new LEMP configuration. The main legacy components that will be replaced for each RF station are the isolator phase-shifter attenuator (I$\Phi$A), the phase and amplitude detectors (PAD), the modulator klystron support unit (MKSU) and the parallel input-output processor (PIOP). A single LEMP LLRF chassis and a solid-state sub-booster (SSSB) will be installed for each RF station. For each sector (8 RF stations), the RF drive unit (RFDU), the SBST klystron and the SBST modulator will also be replaced.

\section{HARDWARE DESIGN}
\subsection{Frequency and Local Oscillator Configuration}
The new system is built on the open-source Marble FPGA carrier and a Zest+ digitizer with a custom RF front-end. A prototype has been deployed and tested at station 26-3.
In the LEMP chassis, all frequencies are phase-locked to the 2856-MHz master oscillator (MO).  
Independent up/down-conversion local oscillators (LO) are synthesized using single-sideband (SSB)
modulation derived from the MO, with separate ADC and DAC clock domains to reduce cross-talk.  
The 2.856-GHz RF is down-converted to 25.5 MHz ($\rm{LO}_{dwn}$ at 2830.5 MHz) and
digitized at 119 MHz.  
The DAC runs at 238 MHz, generating a 93.5-MHz IF up-converted with a
2949.5-MHz $\rm{LO}_{up}$.  
No distributed LO or VCXO is used; all signals remain coherently locked to the MO. Table \ref{Table:LEMP_freqs} shows these key frequencies and their relationships.

\begin{table}[h]
   \centering
   \caption{Frequency relationships}
   \setlength{\tabcolsep}{6pt} 
   \renewcommand{\arraystretch}{1.5} 
   \footnotesize	
   \begin{tabular}{l|c|c|c}
       \hline
       \hline
       \textbf{} & \textbf{Generation} & \textbf{Relationship} & \textbf{$f$ [MHz]}\\
       \hline
           RF & & & 2856\\
           MO & & & 2856\\
           \hline
           $\rm{ADC}_{clk}$ & LMK01801 & $\frac{1}{24}$ MO & 119\\
           $\rm{DAC}_{clk}$ & LMK01801 & $\frac{1}{12}$ MO & 238\\
           $\rm{LO}_{dwn}$ & MO - $\frac{\rm{MO}}{112}$ & & 2830.5\\
           $\rm{LO}_{up}$ & $\rm{LO}_{dwn}$ + $\rm{ADC}_{clk}$ & & 2949.5\\
           $\rm{IF}_{dwn}$ & RF -  $\rm{LO}_{dwn}$ & $\frac{3}{14} \rm{ADC}_{clk}$ & 25.5\\
           $\rm{IF}_{up}$ & Zest+ & $\frac{11}{28} \rm{DAC}_{clk}$ & 93.5\\
       \hline
       \hline
   \end{tabular}
   \label{Table:LEMP_freqs}
\end{table}

\subsection{Analog RF Front-End}

\begin{figure}[ht]
  \centering
  \includegraphics[width=0.9\columnwidth]{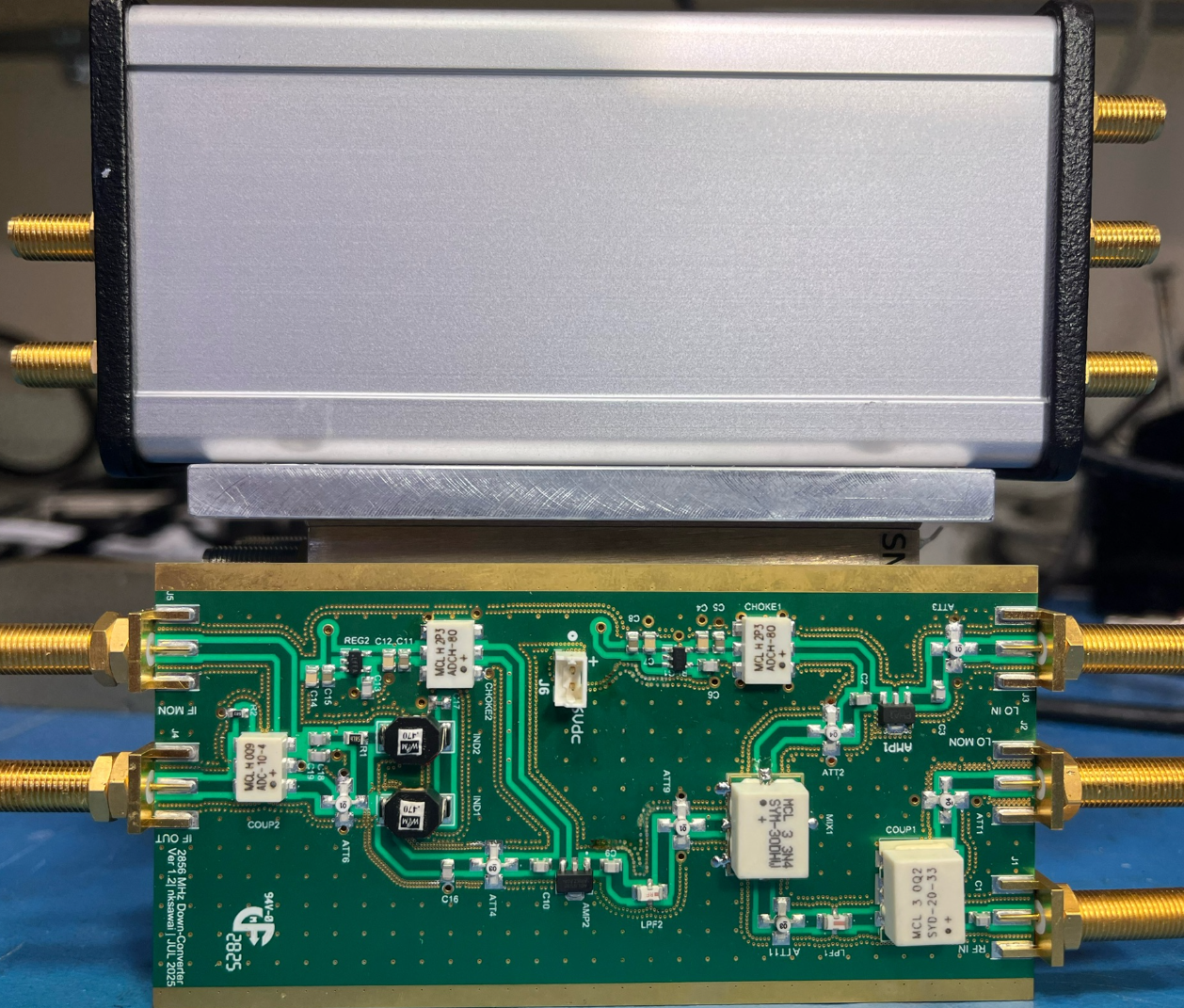}
  \caption{RF front-end downconverter module and enclosure.}
  \label{fig:rf_block}
\end{figure}

The analog front-end down-converts the 2.856-GHz RF to a 25.5-MHz IF using
a mixer-based topology, as shown in Fig.~\ref{fig:rf_block}.  
Each module integrates an RF amplifier, mixer, and low-pass filter optimized
for low noise, high linearity, and stable group delay.  
Independent single-channel modules minimize crosstalk and thermal coupling
compared to multi-channel layouts, improving isolation and reliability. Measured results show the 2$\times$IF spur at 51~MHz is 64~dB below the
fundamental.

On the digitizer board, two input channels were modified from
RF transformer coupling to DC coupling using single-to-differential
amplifiers to enable beam voltage and beam current monitoring.

\subsection{Rear I/O Board}
\begin{figure}[h]
   \centering
   \vspace{-4pt} 
   \includegraphics[width=0.9\columnwidth]{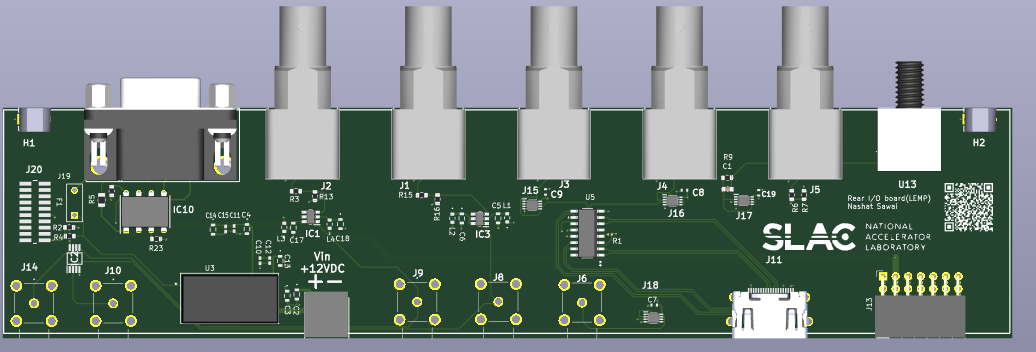}
   \caption{Rear I/O board.}
   \label{fig:rear_io}
   \vspace{-6pt} 
\end{figure}

The Rear I/O board interfaces the LLRF chassis with the klystron system,
replacing the legacy MKSU. It monitors beam voltage and current, attenuates
and buffers signals to the ADC range, and distributes triggers and gate
signals for the modulator and SSSB amplifier through copper and fiber links.
The board manages handshaking and interlocks with the modulator PLC,
ensuring safe operation during fault conditions. Fig.~\ref{fig:rear_io} shows the rear I/O board.

\subsection{Chassis Implementation}

The prototype LLRF chassis integrates all modules for RF generation,
down- and up-conversion, digitization, and interlock control within a single
19-inch enclosure. The layout is optimized for short signal paths, thermal
stability, and ease of maintenance.

\begin{figure}[h]
  \centering
  \includegraphics[width=0.95\columnwidth]{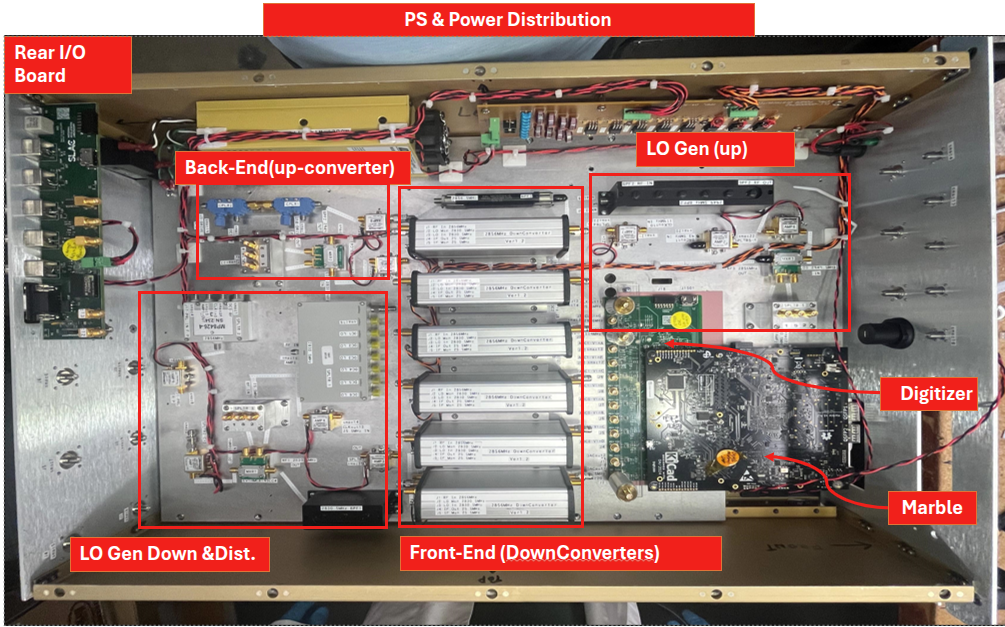}
  \caption{Prototype LEMP LLRF chassis.}
  \label{fig:lemp_chassis}
\end{figure}

As shown in Fig.~\ref{fig:lemp_chassis}, the chassis includes dedicated
boards for rear I/O interfacing, RF down- and up-conversion, LO generation,
and digitization. The \textit{Zest+}~\cite{zest} digitizer board and the \textit{Marble}~\cite{marble} carrier board handle high-speed sampling and FPGA-based processing, while the MMC-driven
\textit{Marble Mailbox} OLED display provides real-time board information such
as IP address, firmware image, and system status.

Power and local distribution circuits are mounted along the side panel under a shielded cover for compactness and EMI isolation. A 14 V DC switched-mode supply feeds a local regulator board that generates 6 V DC and 12 V DC rails. Internal grounding maintains analog–digital separation, minimizing crosstalk and ensuring stable operation during pulsed modes.

\section{Firmware Design}

\begin{figure}[h]
  \centering
  \includegraphics[width=\columnwidth]{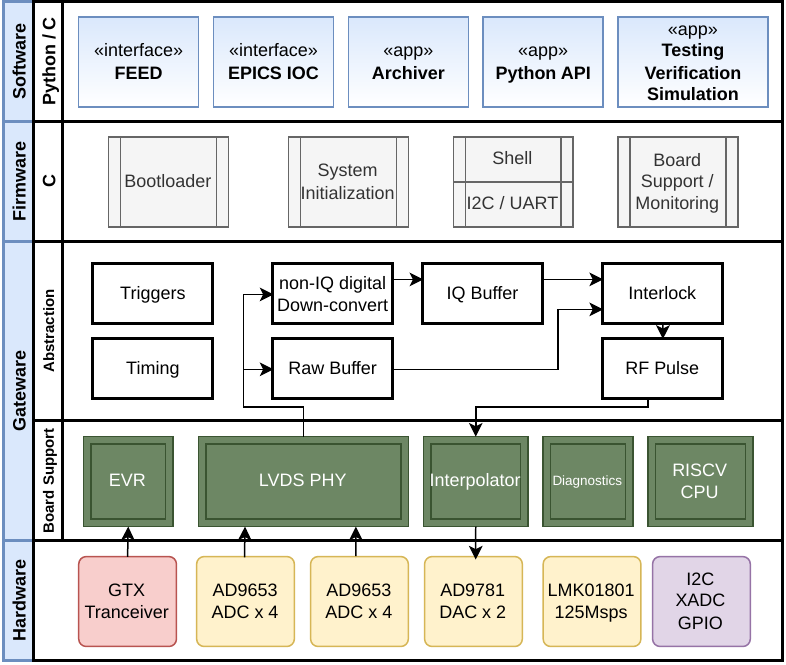} 
  \caption{LLRF system architecture}
  \label{fig:system_architecture}
\end{figure}

The core of the firmware design was developed by the LLRF team at LBNL and is shared between multiple LLRF projects including LBNL's ALS-U AR ~\cite{du2022digital} and the Argonne's AWA project ~\cite{wanming2025}.
It has been modified at SLAC to meet the requirements and functionality of LEMP and LCLS. The system architecture design, which utilizes the BSD-licensed open-source LLRF library Bedrock \cite{bedrock}, is shown in Fig.~\ref{fig:system_architecture}.

The board support layer interacts with the hardware to gather ADC data and timing information, and to transmit the RF output to the DAC. The open-source, size-optimized RISC-V CPU PicoRV32~\cite{riscv} manages boot-time self-initialization and testing, system configuration, and continuous status monitoring. The abstraction layer enables flexible configuration of sampling frequencies, clocks, DSP and EVR settings. System simulation and verification are available with a calibrated cavity model. The packet badger, from Bedrock, provides priority communication to EPICS to access DSP registers, including RF waveforms.

\subsection{DSP structure}
\begin{figure*}[h]
  \centering
  \includegraphics[width=\textwidth]{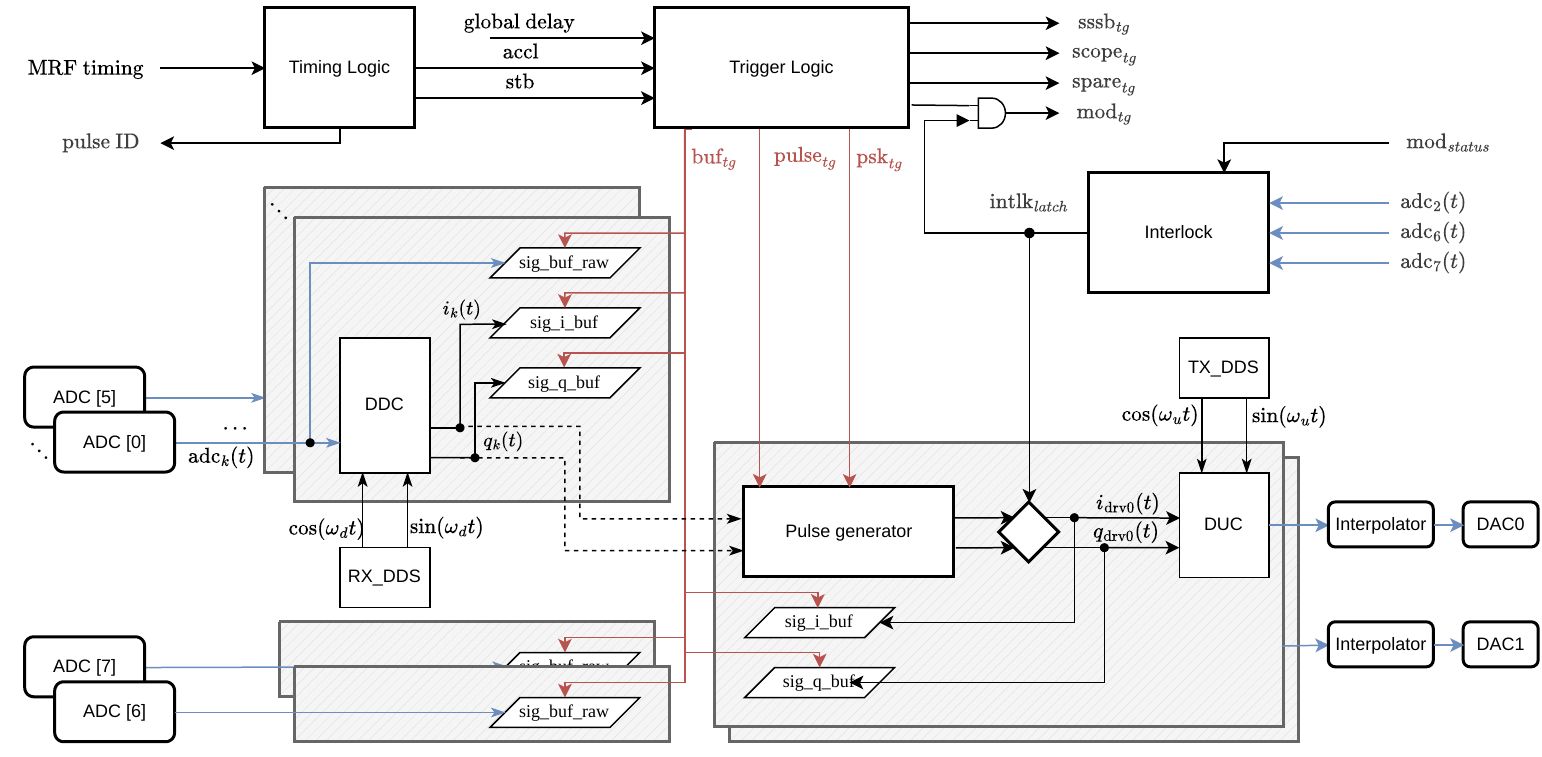} 
  \caption{LLRF DSP Structure}
  \label{fig:dsp}
\end{figure*}

The DSP structure is shown in Fig.~\ref{fig:dsp}. An MRF timing link distributes synchronized events and pulse ID, over a fiber-optic link, that are decoded by an event receiver (EVR). A trigger logic uses the decoded \textit{accelerate} and \textit{standby} events to generate triggers for buffers, RF pulse output, phase shift, solid-state sub-booster (SSSB), and modulator. A global delay, which depends on the operation mode (either accelerate or standby), is applied to all triggers, with each trigger having its own configurable delay and pulse width. An interlock block captures modulator status, klystron reverse, beam voltage, and beam current to generate a trip signal, that disables the modulator trigger and the RF output, if any of the monitored signals exceeds a configurable threshold.

Digital direct synthesis (DDS) generates a pair of sinusoidal signals at 25.5 MHz, or $\frac{3}{14} \rm{ADC}_{clk}$ as shown in Table~\ref{Table:LEMP_freqs}, and with a known starting phase. For high-precision digitization, non-IQ direct digital down-conversion (DDC) is used to avoid aliasing \cite{Doolittle2006DigitalLR}. Digital up-conversion (DUC) is implemented as a double sideband modulator. Both DDC sand 
DUC use the generated sinusoidel signals from the DDS. A final interpolation step takes the output of the DUC to generate a 93.5 MHz signal, which is converted to an analog signal by the DAC. The generated RF pulse has configurable amplitude, phase and width.

Double-buffered buffers are used to store ADC samples for all 8 input channels, and I\&Q data for the 6 RF ADCs and 2 RF DACs. Each buffer size is 1024, which corresponds to $8.6~\mu s$ at the ADC\textsubscript{clk}. Software applications read the I\&Q buffers to generate amplitude and phase plots, along with metadata (pulse ID, ADC min, ADC max) for offline data analysis. 

\section{Software}
\begin{figure}[H]
  \centering
  \includegraphics[width=\columnwidth]{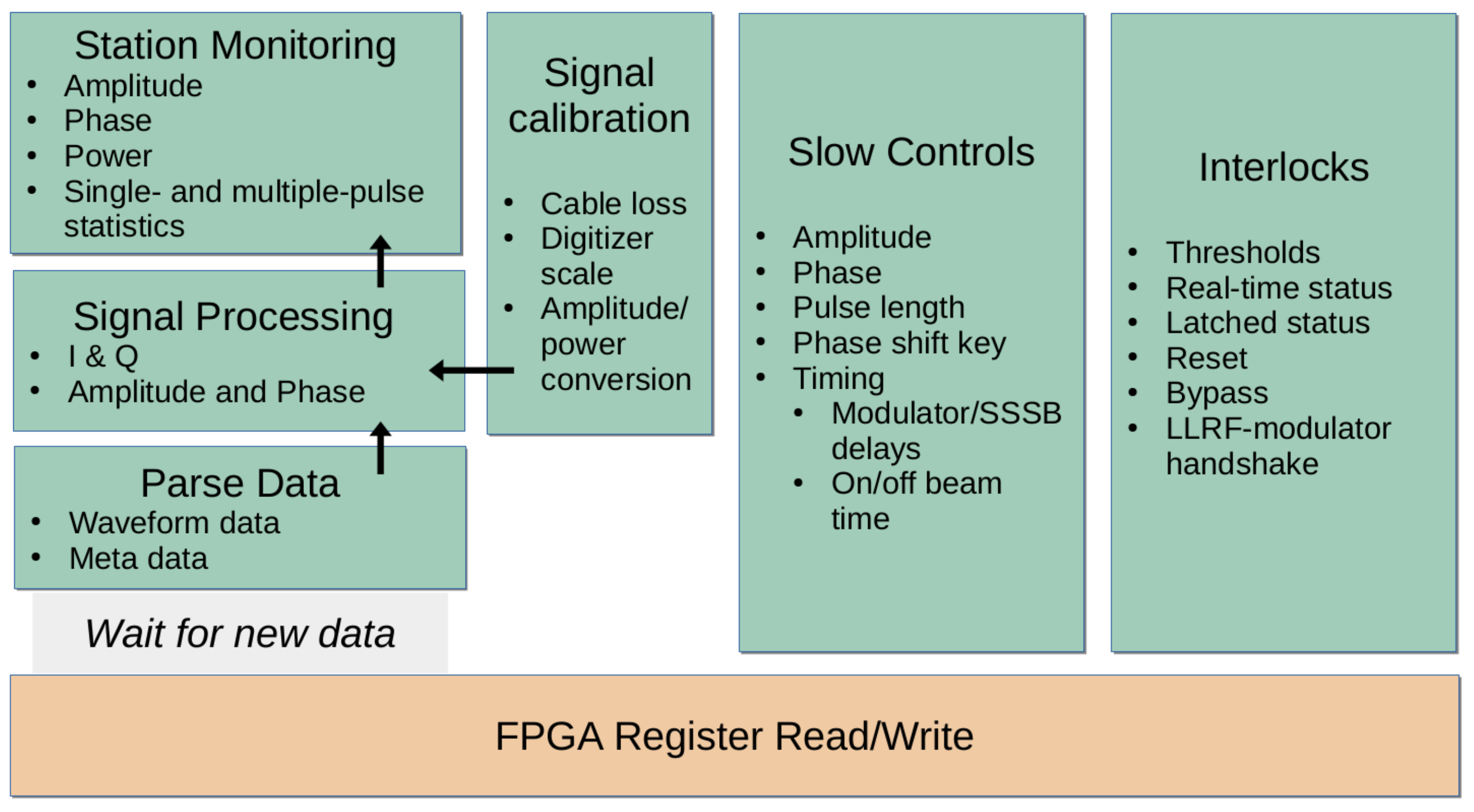}
  \caption{EPICS IOC simplified diagram}
  \label{fig:EPICS}
\end{figure}

An EPICS IOC is used to provide high-level monitoring and control of the
system. It includes software originally developed for the LCLS-II LLRF~\cite{Serrano:IPAC2018-THYGBE3}
system to facilitate communication with the FPGA and for waveform data
processing. New functionality is being added to support the pulsed
operation of the LCLS stations and for integration into the larger LCLS
control system. A simplified diagram of the EPICS IOC is shown in Fig.~\ref{fig:EPICS}

The EPICS IOC runs on a CPU located near the LEMP chassis and communicates
with the FPGA via UDP over a direct Ethernet connection between the CPU
and chassis. A real-time Linux operating system is used to satisfy
requirements for future integration with the LCLS beam data acquisition
system. Another network interface connects the CPU to the larger LCLS
EPICS controls network.

\section{Prototype Test}
The LEMP chassis generates two single-sideband local oscillators: $\rm{LO}_{dwn}=2830.5~\rm{MHz}$ and $\rm{LO}_{up}=2949.5~\rm{MHz}$.
Both are synthesized from the 2856~MHz MO using
a SSB generation scheme to ensure high spectral purity
and phase coherence.The $\rm{LO}_{dwn}$ exhibits a spurious-free dynamic range (SFDR)
greater than $90~\text{dB}$, with MO carrier feedthrough below $-80~\text{dBc}$.
The $\rm{LO}_{up}$ achieves $>81~\text{dB}$ SFDR with carrier feedthrough
below $-80~\text{dBc}$. The RMS additive phase jitter integrated over the $[100~\text{Hz},\,1~\text{MHz}]$ band is 32~fs and 37~fs for $\rm{LO}_{dwn}$ and $\rm{LO}_{up}$, respectively. The RF drive signal at $2.856~\text{GHz}$ after up-conversion demonstrates $84~\text{dB}$ SFDR and RMS additive phase jitter of 38~fs
over the same $[100~\text{Hz},\,1~\text{MHz}]$ band.

\vspace{0.5em}

\begin{table}[!h]
\centering
\caption{Single-sideband LO generation (Down- and Up-Conversion) benchmarks.}
\resizebox{\columnwidth}{!}{%
\begin{tabular}{l|c|c|c}
\hline
\hline
\textbf{Parameter} & \textbf{Down LO} & \textbf{Up LO} & \textbf{Unit} \\
\hline
LO frequency & 2830.5 & 2949.5 & MHz \\
Spurious-free dynamic range (SFDR) & $>90$ & $>81$ & dB \\
MO carrier feedthrough & $\le -80$ & $\le -80$ & dBc \\
RMS additive phase jitter [100Hz, 1MHz] & 32 & 37 & fs \\
\hline
\hline
\end{tabular}%
}
\label{tab:lo_benchmarks}
\end{table}

\vspace{0.3em}

\begin{table}[!h]
\centering
\caption{RF drive signal (Post Up-Conversion, 2.856\,GHz) benchmark.}
\resizebox{\columnwidth}{!}{%
\begin{tabular}{l|c|c}
\hline
\hline
\textbf{Parameter} & \textbf{Value} & \textbf{Unit} \\
\hline
Spurious-free dynamic range (SFDR) & 84 & dB \\
RMS additive phase jitter [100Hz, 1MHz] & 38 & fs \\
\hline
\hline
\end{tabular}%
}
\label{tab:rf_drive}
\end{table}

\begin{figure}[H]
  \centering
  \includegraphics[width=\columnwidth]{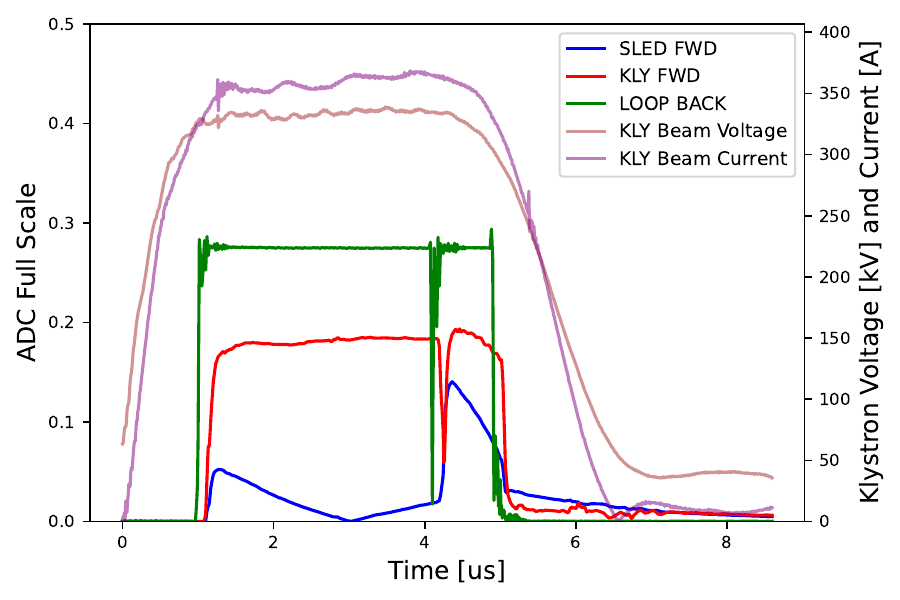} 
  \caption{Waveform acquisition with prototype chassis at station 26-3}
  \label{fig:results}
\end{figure}

The prototype chassis, shown in fig.~\ref{fig:lemp_chassis}, was deployed at klystron station 26-3 (sector 26, klystron \#3) and its main functionality was verified. Fig.~\ref{fig:results} shows the amplitude of the RF signals and the calibrated DC signals of the prototype chassis. According to the standard SLAC NC linac operation, the RF pulse was set to $5~\mu s$ long and the SLED control $180^{\circ}$ phase change occurs after $3~\mu s$ of the start of the RF pulse. These RF parameters are configurable in the LLRF system in order to optimize the RF performance based on local linac considerations.  During the test evaluation phase, the system was set to trigger at the \textit{standby} event, but after successful verification the station was aligned to the \textit{accelerate} event and the phase set point was optimized to increase the beam energy for a user program that allows a beam-based evaluation of the RF performance.  At this time, the LLRF system has been in operation at the klystron station 26-3 for hundreds of hours, though most often in \textit{standby} which enables testing of new development.

\section{Conclusions}
A prototype chassis for the LEMP project has been designed based on previous LLRF projects, tested and deployed at sector 26, klystron \#3, demonstrating stable RF generation, synchronized timing, and < 40 fs phase jitter at 2.856 GHz. After successful verification of the main functionality, the station was aligned to the accelerate event and the phase setpoint was optimized. The system has been in operation for hundreds of hours, mostly in standby mode. Future work includes finalizing the design and building more chassis to upgrade more stations.

\section{Acknoweldgements}
This work is supported by the Office of Science, of the U.S.
Department of Energy, at SLAC under contract number DE-AC02-76SF00515,
and at LBNL under Contract No. DE-AC02-05CH11231.

\ifboolexpr{bool{jacowbiblatex}}%
    {\printbibliography}%
    {%
    
    
} 

\end{document}